# Cryogenic Characterization of the Planck Sorption Cooler System Flight Model

G. Morgante[a1], D. Pearson[b1], F. Melot[c], P. Stassi[c], L. Terenzi[a], P. Wilson[b], B. Hernandez[b], L. Wade[b], A. Gregorio[d], M. Bersanelli[e], C. Butler[a] and N. Mandolesi[a]

[a] *INAF – IASF Bologna,*
  *via P. Gobetti 101, 40129 Bologna, Italy*

[b] *Jet Propulsion Laboratory,*
  *4800 Oak Grove Drive, Pasadena California 91109 USA*

[c] *Laboratoire de Physique Subatomique et de Cosmologie*
  *53 Avenue des Martyrs, 38026 Grenoble Cedex*

[d] *Dipartimento di Fisica, Università degli Studi di Trieste,*
  *via Valerio 2 – I-34127 Trieste, Italy*

[e] *Dipartimento di Fisica, Università degli Studi di Milano,*
  *via Celoria 16, – I20133 Milano, Italy*

  *E-mail*: `morgante@iasfbo.inaf.it, david.p.pearson@jpl.nasa.gov`

ABSTRACT: Two continuous closed-cycle hydrogen Joule-Thomson (J-T) sorption coolers have been fabricated and assembled by the Jet Propulsion Laboratory (JPL) for the European Space Agency (ESA) Planck mission. Each refrigerator has been designed to provide a total of ~1W of cooling power at two instrument interfaces: they directly cool the Planck Low Frequency Instrument (LFI) around 20K while providing a pre-cooling stage for a 4 K J-T mechanical refrigerator for the High Frequency Instrument (HFI). After sub-system level validation at JPL, the cryocoolers have been delivered to ESA in 2005. In this paper we present the results of the cryogenic qualification and test campaigns of the Nominal Unit on the flight model spacecraft performed at the CSL (Centre Spatial de Liège) facilities in 2008. Test results in terms of input power, cooling power, temperature, and temperature fluctuations over the flight allowable ranges for these interfaces are reported and analyzed with respect to mission requirements.

---

[1] Corresponding authors.

# Contents

## 1. Introduction

Planck [1] is a European Space Agency (ESA) mission, whose main objective is to image the temperature anisotropy of the cosmic microwave background (CMB) at high angular resolution. Planck will carry two instruments: the High Frequency Instrument (HFI) [2], based on new generation bolometric detectors, and the Low Frequency Instrument (LFI) [3][4], an array of HEMT technology radiometers. Both the LFI and the HFI instrument sensors need to be cooled to cryogenic temperatures to optimize their signal to noise ratio. The detector cooling system has also to minimize the mechanical vibration to reduce the spurious signal generation on the ultrasensitive detectors.

The LFI radiometers reach the optimal performance point at an operating temperature of 20 K. This temperature is reached through a combination of passive cooling to about 50 K and active cooling using the $H_2$ sorption cooler. The HFI bolometers are cooled to 100 mK through a combination of passive cooling (radiator down to 50 K), the 20 K sorption cooler, a 4.5 K mechanical Joule–Thomson cooler and a Benoit style open cycle helium dilution cooler. The description of the whole cooling chain has been provided earlier [5].

**1.1 Planck Cryochain**

The performance of the ultra-high sensitivity detectors required for the Planck mission is strongly coupled to their temperature. This implies that operating temperatures need to be reached and maintained for the whole mission duration while temperature stability and possible thermal noise due to the cryogenic systems must be controlled to the most accurate level. The requirement for systematic effects minimization is the most important driver for the Planck architecture. This is particularly true for the thermal design, which leads to a cryogenic design that is one of most complex ever for a space mission.

The spacecraft architecture has been optimized to benefit from the favourable thermal conditions of a L2 orbit. The Planck cryo-thermal structure [5] is a combination of passive and active cooling: passive radiators are used as thermal shields and pre-cooling stages; active cryocoolers are used both for instruments cooling and in cascade as pre-coolers for the colder ones.

The global architecture of the Planck cooling system is shown on Figure 1. The 3 active coolers have their source of compressed gas and heat rejection radiators located on the spacecraft service module (SVM) structure (sorption compressor for the $H_2$ cooler, mechanical compressor for the $^4$He JT Cooler, and high-pressure vessels and valves for the dilution cooler). The link between these warm units and the cold end is made via capillaries, with tube-in-tube recuperative counter-flow heat exchangers and thermalization heat exchangers on the pre-cooling stages.

The Planck cryogenic chain can be summarized in the following sequence:

- Solar Array and SVM shield at 300 K to shield the payload from the sun.
- Pre-cooling for all active coolers from 300 K to ~50 K by means of passive radiators in three stages (~150 K, ~100 K, ~50K) [5].
- Cooling to 18–20 K for LFI and pre-cooling for the HFI 4 K cooler with a $H_2$ Joule–Thomson Cooler with sorption compressors (called $H_2$ sorption cooler) [10], [11].
- Cooling to 4 K with a Helium Joule–Thomson cooler with mechanical compressors, as a pre-cooling stage for the dilution refrigerator [7].
- Cooling at 1.6 and 0.1 K with an open loop $^4$He-$^3$He dilution refrigerator [8], [9].

**1.1.1 Passive Cooling**

In the thermal environment of the Earth-Sun Lagrangian L2 orbit, the Sun is the major source of radiation (Earth is practically negligible): the combined action of the sunshield/solar array and the SVM shield protects the rest of the spacecraft from direct solar radiation.

The cold payload is then insulated and shielded from the SVM, where all warm components of instruments and cryocoolers are mounted, by low conductance struts and V-Groove radiators [6]. V-Groove shields are a set of three angled low-emissivity specular surfaces. An open angle of a few degrees between each shield provides each radiator with a view factor to space, allowing an extremely efficient heat rejection. This results in excellent insulation efficiency between objects at different temperatures even with surfaces of moderately low emissivity. In addition, they are very effective in intercepting the spacecraft conductive parasitics (such as loads from mechanical structures, harness, waveguides, cooler pipes) and radiating them to space.

The sequence of three V-Groove shields is used in Planck to lower the temperature of the payload environment down to ~50 K in three steps: approximately 150 K, 100 K and the final 50 K. These three grooves also serve to pre-cool the working fluids of the three coolers in the cryo chain. The coldest radiator, with a surface of only 3-4 m$^2$, is capable of rejecting up to 3 W at 60 K. On the other hand, V-Groove shields impose constraints on the geometry and increase the difficulty of integration.

**1.1.2 Active Cooling**

The objective of the active refrigeration system is to reach and maintain the operating temperatures for the two instruments, while minimizing oscillations and thermal effect that can increase systematic errors level in the mission scientific data.

The Planck H$_2$ Sorption Cooler is the first stage of the active cryogenic chain: its objective is to maintain the LFI to its operating temperature at 20K while providing a pre-cooling stage for the HFI cooling system.

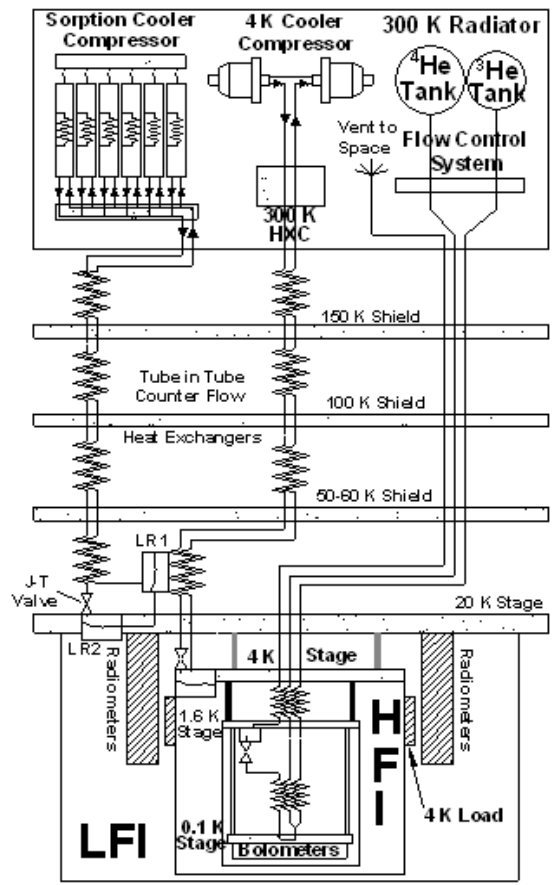

**Figure 1.** Planck cryochain schematic

The dilution refrigerator requires the gas to be pre-cooled at a temperature lower than 10 K. This task is performed by a closed cycle 4 K Joule Thomson (JT) Cooler [7]. This refrigerator uses $^4$He as a fluid, pressurized by a pair of mechanical compressors mounted back-to-back and controlled by low vibration drive electronics with force transducers and a servo feedback loop to minimize the transmitted vibrations.

Part of the heat lift produced by the sorption cooler is used to cool down to 18 K the helium flows of the 4 K and the 0.1 K stages by high-efficiency heat exchangers. These heat exchangers are thermally decoupled from the one used to cool the LFI 20 K plate, for which a larger temperature drop in the exchanger is acceptable.

The Planck lowest temperature stages (1.6 K and 0.1 K) are reached by the 0.1 K Open Cycle Dilution Refrigerator. It exploits a new dilution principle based on friction that does not need gravity to operate [8]. Its cooling power depends on the low gas flow, which allows sufficient gas storage to achieve long mission life [9]. The 0.1 K stage refrigerates the bolometers, thermometers, heaters, and filters. Its temperature is controlled by a closed loop active system. The tubes from and to each stage are attached to form heat exchangers for all circulating fluids in order to minimize thermal losses. The dilution system includes a Joule-Thomson valve, producing a temperature reference stage at 1.6 K for proper insulation of the 0.1 K plate from the radiative and conductive thermal loads coming from the 4 K stage.

## 2. Planck Sorption Cooler Description

Sorption coolers in general are very attractive systems for instruments, detectors and telescopes cooling when a vibration free system is needed. Since the pressurization and evacuation is achieved by simply heating and cooling sequentially sorbent beds, no mechanical vibration is generated. The lack of moving parts like compressors or turbines, increases also the robustness of the system. The only moving parts are the check valves that open and close passively with negligibly small forces, thus essentially creating no vibrations on the spacecraft. This provides excellent reliability and long life. Also, since they employ Joule–Thomson cooling by a simple expansion through a restriction, the cold end can be located remotely from the warm end. Finally, given that the spacecraft's warm end is by design located away (thermally and spatially) from the payload, this allows for excellent flexibility in integration of the cooler to the cold payload (instrument, detectors and telescope mirrors) and the warm spacecraft.

The two Planck sorption coolers are the first continuous closed-cycle sorption coolers to be used for a space mission [10]. They are designed to provide >1 Watt of heat lift at a temperature of <20K using isenthalpic expansion of hydrogen through a Joule-Thompson valve (J-T). More than 80% of this heat lift is used to cool the Low-Frequency Instrument (LFI) down to its operating temperature at 20K [11]. The remaining heat lift is used as a pre-cooling stage for the two cryogenic refrigerators (He J-T cooler to 4K and Dilution cooler to 0.1K) that maintain the High-Frequency Instrument (HFI) at 100mK.

The sorption cooler performs a simple thermodynamic cycle based on hydrogen compression, gas pre-cooling by three passive radiators, further cooling due to the heat recovery by the cold low pressure gas stream, expansion through a J-T expansion valve and evaporation at the cold stage. A schematic of the Planck Sorption Cooler System (SCS) is shown in Figure 4.

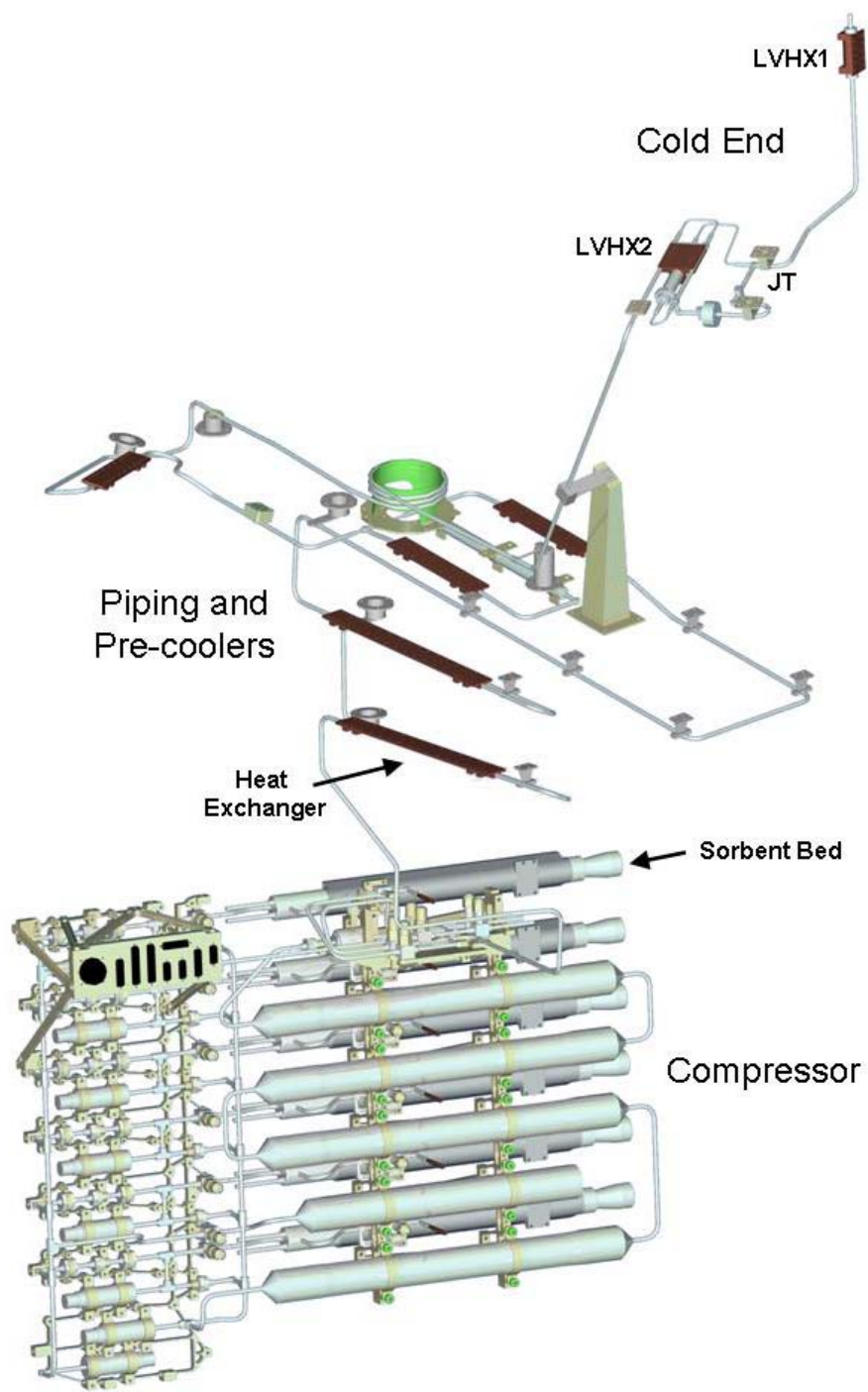

**Figure 2.** 3D view of Sorption Cooler

The principle of operation of the sorption compressor (Figure 3) is based on the properties of a sorption material which can absorb large amounts of hydrogen at relatively low pressure and low temperature, and which can desorb to produce high-pressure hydrogen when heated in a limited volume. The sorbent bed is periodically cycled between heating and cooling cycles, producing high-pressure gas intermittently. In order not to loose excessive amounts of heat during the heating cycle, a heat switch is provided to alternately isolate the sorbent bed from a radiator, which is located on the SVM, during the heating cycle, and to connect it to this radiator thermally during the cooling cycle. As a sorption compressor element (i.e. sorbent bed) is taken through four steps in a cycle (heatup, desorption, cooldown, absorption), it will intake low pressure hydrogen and output high-pressure hydrogen on an intermittent basis. If the high-pressure gas is pre-cooled with radiators to below the inversion temperature and then expanded through a Joule-Thomson expansion orifice (J-T) it will partially liquefy, producing liquid refrigerant at low pressure for sensor systems. Heat evaporates liquid hydrogen, and the low-pressure gaseous hydrogen is re-circulated back to the sorbent for compression.

In order to produce a continuous stream of liquid refrigerant, it is possible to employ several such sorption beds (Figure 3, right) and stagger their phases (like a mechanical gas engine) so that at any given time, one is desorbing high pressure gas while the others are either heating, cooling, or re-absorbing low pressure gas.

For this reason, the Planck compressor assembly shown in Figure 4 is composed of six compressor elements, each connected to both the high pressure and low pressure sides of the plumbing system through check valves, which allow gas flow in a single direction only. The check valves are indicated on the schematic (Figure 4) as single arrows, which indicate the direction of gas flow through them.

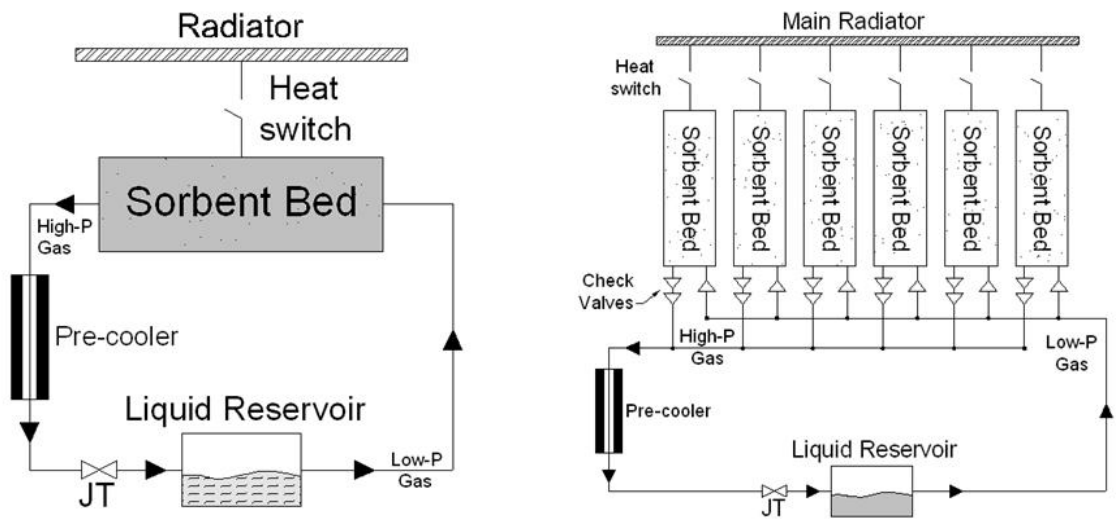

**Figure 3.** Sorption Cooler principle of operation: single bed operation (left panel); full cooler configuration (right)

The high pressure refrigerant then travels from the compressor through a series of heat exchangers and passive radiators, which provide pre-cooling to a temperature below the refrigerant inversion point. Besides the six compressor elements and their check valves, the compressor assembly is also comprised of the high-pressure stabilization tanks and the low pressure stabilization bed. The compressor assembly mounts directly onto the heat rejection radiator.

The pre-cooled gas flow is then expanded through a Joule Thomson (JT) flow restriction valve and the liquid produced is collected in the Liquid Reservoir where it is utilized to provide the required cooling. The instruments heat load partially evaporates the $LH_2$ that is recovered by the compressor through the low pressure line in order to be re-compressed again in a continuous cycle.

The Sorption Cooler System is composed of the Thermo-Mechanical Unit (TMU), the Sorption Cooler Electronics (SCE) and the internal harness. The TMU is the closed fluid circuit that, circulating $H_2$, produces the required cooling. It is composed by the Sorption Cooler Compressor (SCC) and the Piping and Cold End (PACE): a schematic of the TMU is shown in Figure 4. The TMU also includes all the sensors and heaters needed for control and monitoring. The SCE is the hardware/software system that allows TMU operation, control and monitoring. It is the "interface" between the TMU and the Operator.

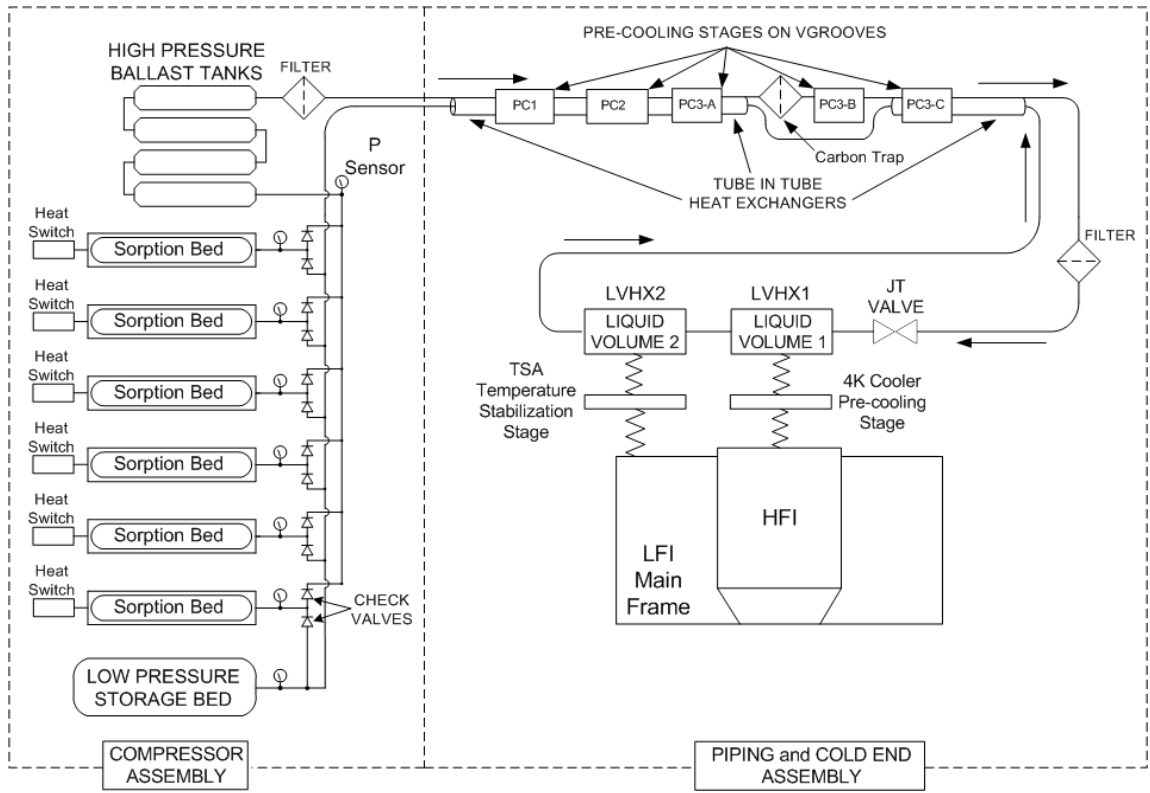

**Figure 4.** Schematic of the Planck SCS TMU showing system components and interface to instruments

## 2.1 Planck Sorption Compressor

The "engine" of the cryocooler is the sorption compressor (Figure 2). It serves two main functions: 1) to produce high-pressure hydrogen gas in the range 3.0 - 4.8 MPa; and 2) to maintain a stable gas recovery rate, which keeps the return pressure, hence the liquid temperature, constant. This is done by the use of compressor elements (or "sorbent beds") [13] whose principle of operation is based on the properties of a unique sorption material that is able to absorb large amounts of hydrogen isothermally at relatively constant pressure and to desorb high-pressure hydrogen when heated to around 200 °C. One gram of the sorbent material, an alloy of Lanthanum, Nickel and Tin ($La_{1.0}Ni_{4.78}Sn_{0.22}$), is able to absorb up to 140 scc of $H_2$ gas at saturation. Each bed in the compressor contains approximately 600 g of hydride powder.

Heating of the sorbent material is accomplished by electrical resistance heaters while the cooling is achieved by thermally connecting the compressor element to a radiator sized to reject the cooler input power at 270 K ±10 K. Six compressor elements are required for the compressor to operate cyclically (Figure 3, right panel, and Figure 4). At any moment one bed is releasing gas (desorption) high pressure, three are absorbing gas to maintain the vapour pressure constant, while the other two beds are being heated and cooled in preparation for desorption and absorption respectively. The ability of the compressor to maintain the vapour pressure of the liquid constant is determined by the absorption properties of the sorbent material. As a compressor element fills with hydrogen gas, the pressure in the return line will rise slightly and this is the main source of temperature fluctuations at the two liquid-vapour heat exchangers (LVHXs). The cycle time of the compressor can range from 550 to 1200 seconds in nominal operations and is adjusted according to cooler performance and requirements.

As each compressor element undergoes the cyclic heating and cooling, a gas-gap heat switch is used to couple or decouple the compressor element to the radiator depending on its state [14]. The heat switches use a sorbent material that when heated releases gas to turn the switch "ON" and when cooled reabsorbs the gas to isolate the element. During the heat-up and desorption cycles the heat switch is "OFF", while during the cooldown and absorption cycles the heat switches are "ON".

The compressor also includes four 1-liter tanks on the high-pressure side (HPST) as shown in Figure 4. These tanks serve as a gas ballast to smooth mass flow variations due to the desorbing compressor elements. On the low pressure side of the compressor is a low pressure storage bed (LPSB) that stores hydrogen gas when the cryocooler is not operating to keep the system pressure below 1 Bar. Additionally, the LPSB stores gas that is evolved as the cooler ages. Two heaters are mounted to the LPSB. One is used in nominal operation to control the gas concentration in the compressor elements, while the second is used when the cooler is started to move gas from the LPSB to the HPST. Check valves direct flow out of the compressor elements into the HPST and control flow from the low pressure manifold and the LPSB back into the absorbing beds.

**2.2 Piping Assembly and Cold End (PACE)**

The Piping Assembly and Cold End comprises the two main parts of the PACE (Figure 2). The Piping Assembly consists of a tube-in-tube heat exchanger and three pre-cooler interfaces. This subsystem serves to pre-cool the high-pressure gas stream to below 60 K to produce the required cooling power. The three pre-coolers heat exchangers attach to V-groove radiator panels with nominal temperatures of 150 K (PC1), 100 K (PC2), and 50 K (PC3). For PC3, three stages are implemented to optimise heat exchange with the radiator. A carbon cold trap is also located on the coldest radiator to remove condensable contaminants from the high pressure gas stream.

As shown in Figure 4, the Cold End, as the second assembly, consists of the Joule-Thomson expander, the two liquid-vapour heat exchangers and a Temperature Stabilization Assembly (TSA) for the LFI instrument. The JT expander is selected to produce a flow up to 6.5 mg/s for an input pressure of 4.8 MPa. The first liquid-vapour heat exchanger, LVHX1, attaches to the HFI instrument. It is designed to provide a temperature lower than 19 K with 190 mW of cooling power. The second LVHX, attaches to the LFI instrument to provide a temperature less than 22.5 K and >646 mW of cooling power. At the interface of LVHX2 and LFI, a copper block is designated as the Temperature Stabilization Assembly. Two stainless steel strips are sandwiched in between to define the conductance between the TSA and LVHX2. This arrangement allows active temperature control at the interface using a PID algorithm. 150 mW are allocated for the TSA for implementation of this temperature control scheme. In addition, the high-pressure gas stream exchanges heat with LVHX2 to pre-cool the gas and maintain its temperature constant before passing through the JT expander. Other elements of the cold-end include a tube-in-tube heat exchanger that joins the last pre-cooler to the cold-end, and a particle filter that protects the JT expander.

**2.3 SCS operations**

The Sorption Cooler performance and lifetime depends primarily on the temperatures of the two main interfaces and the heat load from the two instruments. The two main interfaces are the warm radiator (WR) and the final pre-cooling stage on V-Groove 3.

In Figure 5 a schematic view of the main thermal interfaces is shown. The operational concept of the system is relatively simple. Cooling power is function of the pre-cooling stage (basically the third V-Groove) temperature and the mass flow rate across the J-T throttle valve. A lower V-Groove 3 temperature provides more heat lift capability.

The cold end temperature is a function of the absorbing pressure in the hydride beds. The absorbing pressure is principally a function of the warm radiator temperature.

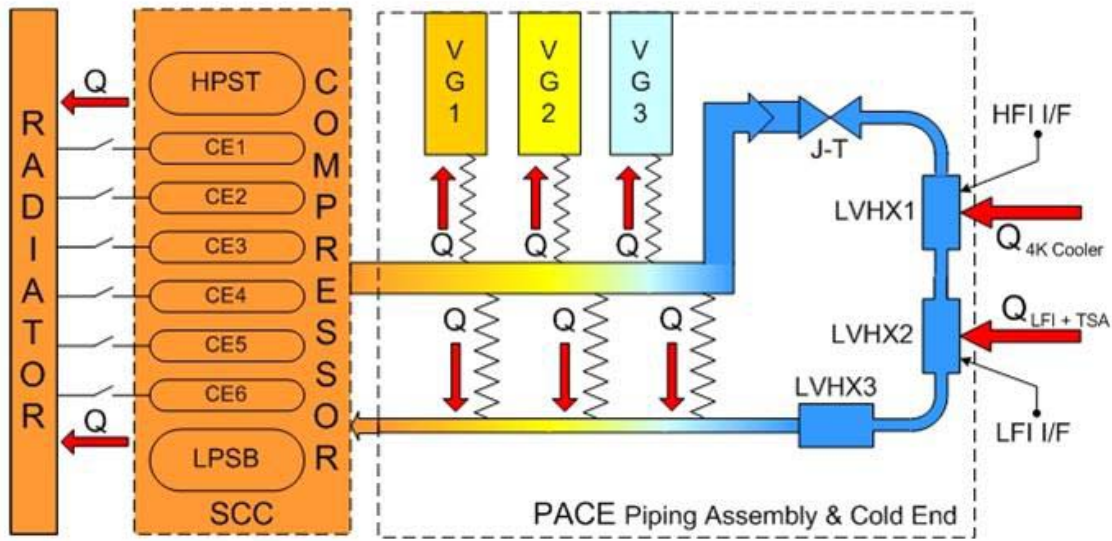

**Figure 5.** Main thermal interfaces

The approach is to operate the Sorption Cooler such that science performance is maximised while meeting the requirements. Since the lifetime of the two Sorption Coolers is critical, the system is run in such a way to maximise lifetime. This is done essentially by "throttling" the cooler using two main parameters to provide only the required cooling. These two main parameters are:

1) cycle time

2) input power

As the hydride of the Sorption Cooler ages, it will be necessary to decrease the cycle time and increase the input power to keep the cooler operating within temperature fluctuation and heat lift requirements. The power has to be adjusted within the allocated range from beginning-of-life to the end-of-life operations. The cycle time will be set at the maximum value at the beginning of operations and reduced regularly to cope with system degradation. Its value has to be properly set in order to avoid harmonics of the 60-second spin cycle of the Planck spacecraft that, being modulated as the sky, would show up in the data as signal.

**2.4 SCS requirements**

The key requirements of the Planck sorption cooler are summarized below:

- Provide ~1W total heat lift at instrument interfaces using a <60 K pre-cooling temperature at the coldest V-groove radiator on the Planck spacecraft
- Maintain the following instrument interfaces temperatures:
    - LFI @ < 22.5 K [80% of total heat lift]
    - HFI @ < 19.02 K [20% of total heat lift]
- Temperature stability (over cooler cycle time):
    - 450 mK, max. to min. at HFI interface
    - 100 mK, max. to min. at LFI Interface
- Input power consumption 470 W (end of life; excluding electronics)
- Operational lifetime: a total of 18 months for both coolers, 15 months for two sky surveys plus 3 months for calibration and performance verification in flight

| TMU Spec | Requirement Value |
|---|---|
| **Cold End Temperature** | 17.5 K < LVHX1 < 19.02 K<br>17.5 K < LVHX2 < 22.50 K |
| **Cooling Power** | Cooling power @ LVHX1 > 190 mW<br>Cooling power @ LVHX2 > 646 mW<br>TSA dissipation = 150 mW<br>Total Cooling Power > 986 mW |
| **Input Power** | TMU Input power < 426 W @ BOL<br>TMU Input power < 470 W @ EOL |
| **Cold End Temperature Fluctuations** | $\Delta$T @ LVHX1 < 450 mK pp<br>$\Delta$T @ LVHX2 < 100 mK pp |

**Table 1.** Primary verification criteria for SCS TMU performance

In Table 1 are summarized the SCS requirements: they are the primary reference values to evaluate the cooler performance.

### 2.4.1 Planck SCS Ground Characterization at System Level

Planck SCS units were tested at JPL before delivery to ESA [12]. Both units have been tested versus main interfaces flight allowable ranges in a dedicated cryofacility simulating the spacecraft.

After delivery to ESA, the two units have been checked on the flight spacecraft to verify their compliancy to the requirements in conditions representative of flight. Since there were orientation constraints for SCS operations on ground, the two units could not be tested in the same campaign. Two test campaigns, called PFM1 and PFM2 each dedicated to one unit, were performed at the CSL (Centre Spatial de Liège) facilities in 2006 and in 2008. The first campaign was used to test the redundant unit (FM1). During that test the main interfaces could still be artificially controlled to cover all range of flight allowable. Results of this campaign are summarized in Table 2.

In summary the PFM1 test campaign confirmed the functionality of the SCS at hardware, software and operating level: the SCS performed as expected and, in few cases, even better. Parameters like Cold End temperatures and fluctuations, heat lift, input power were in most cases compliant to the requirements and comparable to previous ground tests. Due to on-board software issues it was not possible to activate the temperature control on the TSA stage during PFM1 test: for this reason compliancy of the TSA to the 100 mK peak-to-peak requirement could not be verified. In Table 2 are reported only the raw temperature fluctuations measured at the LVHX2. Temperature stability at the HFI interface, LVHX1, was within requirement except for the Reference case. These excess fluctuations are believed to be caused by gravitationally induced plugs due to the particular orientation of the cold end with respect to gravity during ground test. A waiver to this requirement was agreed by the instruments and ESA management.

| Test case | WR Average T [K] | PC3C T [K] | LVHX1 T [K] | LVHX1 ΔT [mK] | LVHX2 T [K] | LVHX2 ΔT [mK] | Heat Lift [mW] | Input Power [W] | Cycle Time [s] |
|---|---|---|---|---|---|---|---|---|---|
| SCS Requirements | | | <19.02 | <450 | <22.50 | <100 | > 986 | <470@EOL | |
| Cold SCC Thermal Balance | 270.5K | 45 | 17,2 | 422 | 17,3 | 556 | 1100 ± 50 | 297 | 940 |
| Reference SCC Thermal Balance | 282.6K | 60 | 18,4 | 497 | 18,6 | 600 | 1050 ± 50 | 388 | 667 |
| Hot SCC Thermal Balance | 276.9K | 60 | 18,0 | 307 | 18,8 | 325 | 1100 ± 50 | 458 | 482 |

Table 2. PFM1 Test results for Redundant unit

## 3. PFM2 Nominal Unit Test

### 3.1 Test Objective

The Planck sorption cooler Nominal unit (FM2) was tested between June and August 2008, for a total of 41 days. The main goal of the PFM2 test campaign was the functional validation of the Sorption Cooler Nominal Unit (SCS-N) on the S/C with the Planck payload in full flight representative conditions (Warm Radiator and V-grooves at Flight Nominal Temperature). Two main thermal cases were run: Warm Radiator cold; and Warm Radiator hot. For the cold case, that represents the beginning-of-life conditions, the SCS is tuned to produce the heat lift necessary for the instruments and its operation, subject to the constraint of maximizing SCS lifetime. For the hot case the SCS is tuned to produce its maximum power, 470 W, which will simulate end-of-life conditions. For each case the SCS must meet its requirements, Table 1. In addition, many other minor tests were performed, mainly on operational and safety issues, that are not reported here. We present here only the relevant cryogenic results.

Lifetime of the sorption cooler system was more accurately assessed with the results of the current testing. With the determination of the $3^{rd}$ V-Groove temperature and the LFI instrument load, a more refined estimate of the lifetime was made. With these conditions as baseline, the nominal cooler will be capable of operating for 15.5 months, while the redundant can operate for 13.5 months. Combining both coolers lifetime allows a total of 29 months that exceeds the minimum mission requirement of 18 months (see Par. 2.4).

## 3.2 Test configuration

The PFM2 test configuration was equivalent to the final flight configuration. Warm Radiator was controlled as in flight and the cryo-facility radiative environment was close to deep sky temperature as seen from L2, on the order of 5-10 Kelvin. In these conditions the radiators could reach their natural balance point with respect to cooler dissipation at the different stages.

## 3.3 Test Results

### 3.3.1 SCS startup and cooldown

The SCS was started on 27 June. Liquid hydrogen was produced 200 hours later and the nominal mode was entered 20 hours after liquid production.

Objective of this test is to start the SCS-N and take it into Run Mode and Nominal Operations. The startup and cooldown process were performed as expected. Cooldown was within the predictions of the Spacecraft thermal model. During the cooldown the SCS performed without any issues. The entire cool down process is shown in Figure 6.

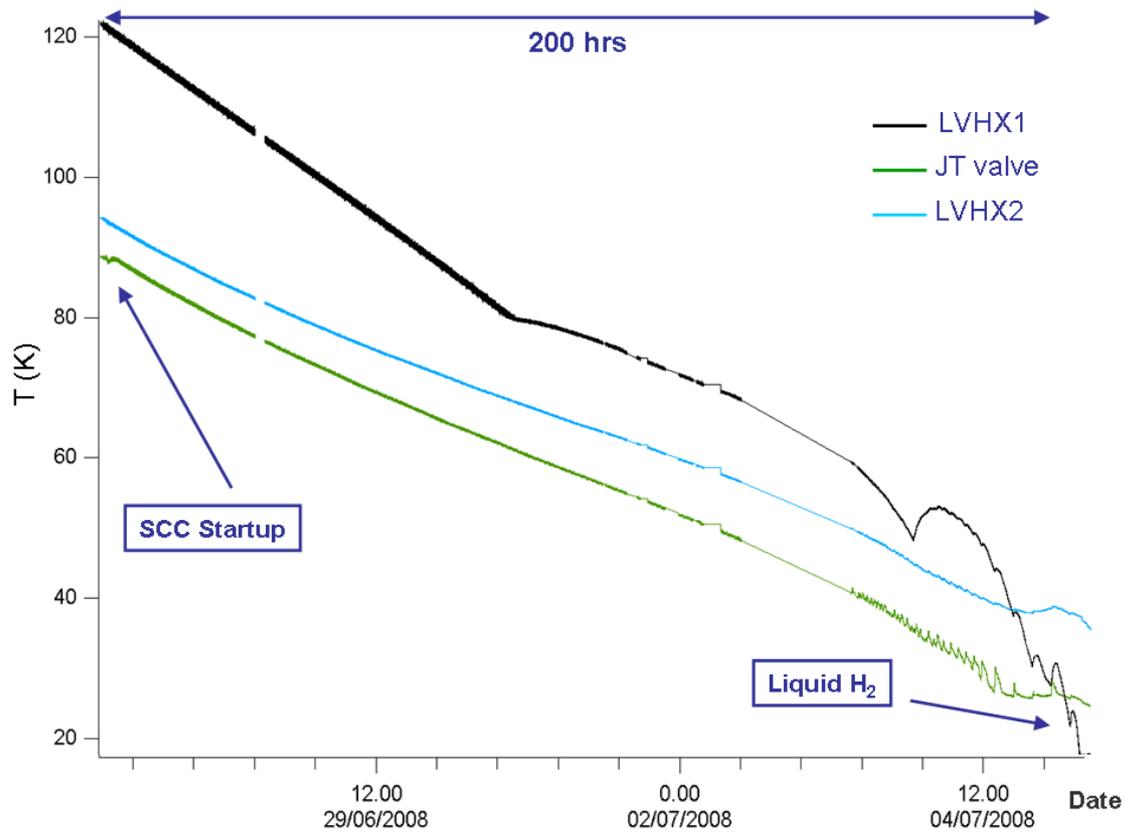

**Figure 6.** Cooldown of the SCS cold-end. Total time from SCS start-up to the production of liquid was 200 hours. Temperature traces are: black LVHX1; blue LVHX2; green Joule-Thomson valve (K).

A zoomed view of first liquid hydrogen production and of transition into Normal Mode is shown in Figure 7.

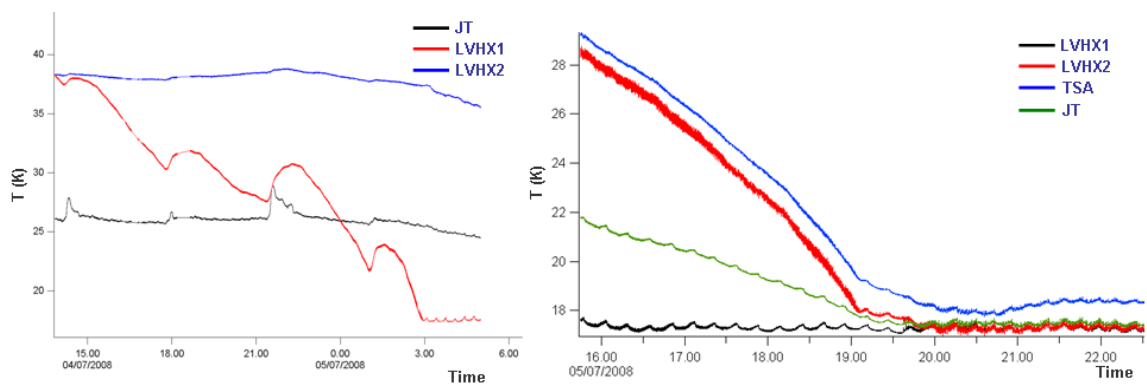

**Figure 7.** On left panel: 20K reached and liquid produced in LVHX1 (red line). On the right side is shown the transition in Normal Mode, when the whole Cold End reaches 20K

### 3.3.2 Cold Thermal Balance, Beginning of life conditions

The objective of this test was to adjust the SCS operational parameters for nominal cooler performance in order to check performance and operation settings in BOL conditions. The results of this test provide a fundamental reference for cooler operations in flight in similar boundary conditions. The two primary interfaces, V-groove three and warm radiator temperatures were 47 and 270 K, respectively. The SCS ran almost 10 days under these conditions. Table 3 summarizes the cooler performance results for this case.

### 3.3.2.1 Temperature and temperature fluctuations

Temperature data (from top to bottom) for the TSA, LVHX2, and LVHX1 are shown in Figure 8. The TSA stage is controlled to a set point of 18.7 K, with fluctuations of 120 mK. These fluctuations are greater than the requirement of 100 mK. For LVHX1 the temperature is 17.09 while fluctuations are about 550 mK, again greater than the requirement. In addition, the power to control the TSA is ~200 mW that exceeds the 150 mW requirement.

Each of these requirement excesses are attributable to gravitationally induced plug flow, which were to be expected for the cold-end orientation for this test (the excessive TSA power is mainly due to the use of a constant set point of 18.7 K, but some of the excess power is due to the plug flow). Finally, a cooler cycle modulation is clearly present in the data, i.e. due to differing performances of the individual compressor elements, the temperature of the cold-end varies over 6 bed cycle.

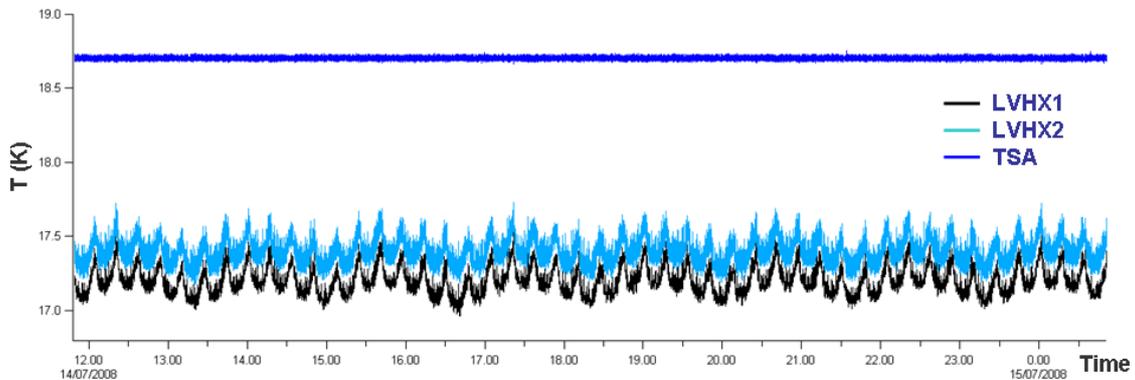

**Figure 8** Cold Case cold-end temperatures. The dark blue trace is the TSA stage, the lighter blue is LVHX2, and the black is LVHX1.

### 3.3.2.2 SCS Heat Lift Measurement

The objective of this test is to measure the cooling power produced by the SCS in the Cold Thermal Balance Case. This is a fundamental verification of cooler functional performance and LFI thermal behaviour: it allows not only to measure SCS performance in terms of heat lift but also to provide an indirect estimation of the LFI passive dissipation (parasitics).

Heat lift of the sorption cooler was evaluated by determining the heat lift excess using the TSA heater, and by estimating the instrument loads. The excess heat lift was measured to be 355 mW, where 200 mW is for the TSA. The LFI load was estimated by use of the temperature difference between the TSA and the LVHX2, and the JPL measured thermal resistance between these two stages. This gave about 670 mW +/- 50 from LFI. HFI was assumed to be 100 mW, with the same uncertainty of +/- 50 mW. Thus the total heat lift was 1125 mW.

Calculating the heat lift from the SCS working pressures and PC3C temperatures gives a value of 1155 mW. This compares well to the measured 1125 mW. Thus, to experimental uncertainties, the SCS was performing nominally.

| TMU Spec | Cold Case Results | Requirement Value |
|---|---|---|
| Cold End Temperature | 17.09 K LVHX1<br>18.7 K TSA | 17.5 K < LVHX1 < 19.02 K<br>17.5 K < LVHX2 < 22.50 K |
| Cooling Power | 1125 ± 75 mW | Cooling power @ LVHX1 > 190 mW<br>Cooling power @ LVHX2 > 646 mW |
| Input Power | 304 W | TMU Input power < 426 W @ BOL |
| Cold End Temperature Fluctuations | 550 mK LVHX1<br>120 mK TSA | ΔT @ LVHX1 < 450 mK<br>ΔT @ LVHX2 < 100 mK |

**Table 3.** Cold Case results summary

### 3.3.3 Hot Thermal Balance, End of life conditions

Objective of this test was to verify SCS performance in Hot EOL conditions and its impact on SVM thermal balance. Major thermal interfaces were taken up to worst case temperatures in order to check PPLM functionality in these conditions. SCS shall be verified in worst case, end of life, maximum dissipation case: 470 W.

The SCS performance are summarized in Table 4. **Hot Case results summary** The total input power was 470 W, maximum allowed value, when the cooler was in normal mode. In order to replicate the full end-of-life conditions, the cycle-time was 525 s. With this input power the warm radiator reached a temperature of ~273 K. This temperature was not stable for the test period. As a consequence the cold-end temperature was observed to drift. Likewise, the 3$^{rd}$ V-groove drifted from about 46.9 to 47.7 K over the test period. Since the SCS was run with a heat-lift excess of about 0.5 W, performance was not impacted.

#### 3.3.3.1 Temperature and temperature fluctuations

Fluctuations for the hot-case were ~350 mK peak-to-peak for LVHX1 and 60 mK peak-to-peak for the TSA stage. Here the SCS requirements are met. This is in contrast to the cold-case where the requirements were not met for the two instrument interfaces. The difference is most likely due to the differing heat lift conditions of the two cases, which leads to different two-phase flow environments, i.e. different locations for the final liquid interfaces.

| TMU Spec | Cold Case Results | Requirement Value |
|---|---|---|
| Cold End Temperature | 17.47 K LVHX1<br>18.7 K TSA | 17.5 K < LVHX1 < 19.02 K<br>17.5 K < LVHX2 < 22.50 K |
| Cooling Power | Not measured | Cooling power @ LVHX1 > 190 mW<br>Cooling power @ LVHX2 > 646 mW |
| Input Power | 470 W | TMU Input power = 470 W @ EOL |
| Cold End Temperature Fluctuations | 350 mK LVHX1<br>60 mK TSA | ΔT @ LVHX1 < 450 mK<br>ΔT @ LVHX2 < 100 mK |

**Table 4.** Hot Case results summary

## 3.4 Conclusions

During PFM2 test campaign the Planck SCS Nominal Unit was characterised in flight representative conditions. Two main thermal configurations were verified: beginning-of-life; and end-of-life power conditions. These two cases correspond to input powers of 304 and 470 W, and hot radiator interfaces of 270 and 273 K for the cold and hot cases respectively. For the cold case the last pre-cooling stage (PC3C or V-groove 3 temperature) was 47 K and 48 K for the hot case. Cooler performance is determined by these two interfaces. For the testing, the sorption cooler met all of its 4 main requirements- cold-end temperature and fluctuations, heat lift or cooling power, input power- except for temperature fluctuations.

The observed fluctuations on the LVHX1 were 550 mK versus a requirement of 450mK. These excessive fluctuations were identified as being due to two-phase plug flow events at a period of about 20 s. The temperature fluctuation requirement was also not met in the JPL sub-system testing nor the ESA PFM1 test for the same reason.

Early instrument reports indicated that the measured high fluctuation levels would not impact instrument performance due to the high frequency of the two-phase flow events.
Instrument heat loads were estimated to be 670 mW from LFI, with a less than 30 mW contribution from HFI. The TSA power consumed was 200 mW. An additional 90 mW was allocated for parasitics and operations margin. There is a relatively large uncertainty in the estimated instrument loads, but the total number is consistent with the cooling power measurement performed during the PFM2 testing to a level of +/- 50 mW.

Finally, the input power used for the testing was 304 W. This is well below the requirement, and lifetime predictions show that input power will not be the limiting factor for a 47 K V-groove temperature.

The estimated lifetime of the two flight models is 15.5 months for FM2 and 13.5 for FM1 based on the PFM2 testing conditions (i.e. V-groove 3 temperature = 47 K, and total nominal heat load of 1060 mW) and assuming that the cycle-time can be increased due to the absence in flight of the two-phase flow events. These estimates have an uncertainty of order 0.5 month. The end of life temperature of the cooler cold end is expected to be higher by at most 0.5 K.
In summary the measured performance is satisfactory for meeting the ambitious goals of the Planck mission.

## Acknowledgments

The Planck Sorption Cooler System has been designed, fabricated, assembled and qualified at the Jet Propulsion Laboratory, California Institute of Technology, under a contract with NASA. Planck is a project of the European Space Agency with instruments funded by ESA member states, and with special contributions from Denmark and NASA (USA). The Italian contribution to Planck SCS has been supported by the Italian Space Agency (ASI).